\documentclass[sigconf]{acmart}

\usepackage{booktabs} 
\usepackage{amsmath}
\usepackage{graphicx}
\usepackage{multirow}
\usepackage{natbib}
\usepackage{comment}
\usepackage{enumitem}
\usepackage{multirow}
\usepackage{tablefootnote}
\usepackage{subcaption}
\usepackage{xurl}

\setcopyright{rightsretained}

\renewcommand\footnotetextcopyrightpermission[1]{}

\acmDOI{}

\acmISBN{}

\acmConference[PCSC2026]{Philippine Computing Science Congress}{April 2026}{Davao, Philippines}
\acmYear{2026}
\copyrightyear{2026}

\acmArticle{}
\acmPrice{}

\editor{}
\editor{}
\editor{}

\begin{document}
\title{Towards More Empathic Programming Environments}
\subtitle{An Experimental Empathic AI-Enhanced IDE}

\author{Justin Rainier Go}
\email{justin\_rainier\_go@dlsu.edu.ph}
\orcid{0009-0008-2426-4517}
\affiliation{
    \institution{De~La~Salle~University}
    \city{Manila}
    \country{Philippines}
}

\author{Roemer Gabriel Caliboso}
\email{roemer\_caliboso@dlsu.edu.ph}
\affiliation{
    \institution{De~La~Salle~University}
    \city{Manila}
    \country{Philippines}
}

\author{Kurt Christian Andaya}
\email{kurt\_andaya@dlsu.edu.ph}
\affiliation{
    \institution{De~La~Salle~University}
    \city{Manila}
    \country{Philippines}
}

\author{Aaron Daniel Go}
\email{aaron\_daniel\_go@dlsu.edu.ph}
\affiliation{
    \institution{De~La~Salle~University}
    \city{Manila}
    \country{Philippines}
}

\author{Jocelynn Cu}
\email{jocelynn.cu@dlsu.edu.ph}
\affiliation{
    \institution{De~La~Salle~University}
    \city{Manila}
    \country{Philippines}
}

\renewcommand{\shortauthors}{Go et al.}

\begin{abstract}
    As generative AI becomes integral to software development, the risk of over-reliance and diminished critical thinking grows. This study introduces ``Ceci,'' our Caring Empathic C IDE designed to support novice programmers by prioritizing learning and emotional support over direct code generation. The researchers conducted a comparative pilot study between Ceci and VSCode + ChatGPT \cite{ds24,s25}. Participants completed a coding task and were evaluated using the NASA-TLX workload assessment and a post-test usability survey. Although the sample size was small $(n=11)$, results show that there is no significant difference in perceived effectiveness, learning and workload between the Experimental Ceci group and the Control group, though Ceci users reported significantly greater perceived helpfulness in error correction ($p = 0.0220$). These findings suggest that empathic responses may not be sufficient on their own to enhance the learner's outcomes, perceptions, or reduce workload. Overall, this study provides a foundational framework for future research. Such research should explore larger sample sizes, diverse programming tasks, and additional empathic features to better understand the potential of empathic programming environments in supporting novice programmers; they must also ensure that the empathic features are well-integrated in the user interface.
\end{abstract}

%
%



\maketitle


\section{Introduction}

\subsection{Background of the Study}

The field of artificial intelligence (AI) is rapidly permeating all aspects of software development.
Coding assistants and large language models (LLMs) like Gemini, Copilot, and ChatGPT are now being used not just for generating code but also for modernizing legacy codebases through code migration, ensuring the security of code bases through API misuse detection, and enforcement of best practices through code review \cite{mmlsgs25,mrwam25,vsbzliclaptmj24}.
Despite this, AI-assisted software development is still limited by its inability to generate entire codebases from scratch, refactor existing code while ensuring it passes its own generated tests, and can even hinder innovation in the field \cite{esc25,gllm25,plb23}.
These rapid advances in AI for professional development raise the question of how similar tools might support learning and teaching in programming education.

As artificial intelligence becomes an increasingly integral part of our world, its potential to create personalized learning experiences is a significant benefit \cite{lgm23}.
In programming education specifically, students already use AI tools to debug code, understand concepts, and generate frameworks \cite{gvrv24}.
However, simply providing technical assistance is not enough.
While empathy is a well-established positive trait in human educators, its integration into AI (and the numerous effects of doing so) remains a critical area of investigation \cite{ack22}.


Given that the debugging process is a well-documented source of frustration for novice programmers~\cite{bd15}, there is an opportunity to explore how AI systems can go beyond correctness and efficiency to also support learners emotionally. This paper addresses that challenge by exploring the synergy between empathic systems and programming education through the design of an empathic debugger.

Furthermore, unlike prior work that primarily focuses on improving code generation or debugging accuracy, this study investigates the role of affective scaffolding in programming environments. Rather than optimizing correctness, the goal is to examine whether emotional support can influence cognitive load, frustration, and perceived learning outcomes.

By shifting the focus from performance-centric AI assistance toward learning-oriented AI design, this work contributes to the emerging area of empathic computing in software engineering education.

\subsection{Research Questions}

The main problem of this study is to determine how the use of an empathic AI chatbot interface during debugging or compilation affects novice programmers' frustration levels and learning outcomes in introductory programming.

Specifically, the study aims to answer the following questions:

\begin{enumerate}
	\item What are novice programmers' perceptions of the effectiveness of an empathic AI chatbot during coding and code compilation tasks?
	\item To what extent does interaction with an empathic AI chatbot enhance novice programmers' understanding of fundamental programming concepts?
	\item How effective is the empathic AI chatbot in reducing cognitive load and frustration among novice programmers?
\end{enumerate}


\section{Review of Related Literature}

\subsection{Frustrations for Novice Programmers}

Research indicates that learning to program is a deeply emotional experience for novices, with frustration often cited as a critical barrier to success. It is easier for these students to get stuck and reach states of confusion and frustration than it is going back to being engaged.

Categories of frustration as categorized by \citet{fp15} stems from plenty of factors intrinsic to computer science such as a misunderstanding of the mapping between behavior and cause, learning new programming technologies, adjusting to a new project environment, unavailability of quality and beginner-friendly doc- umentation, as well of some psychologically intrinsic categories such as fear of failure and internal pressure. Some participants also noted some external variables which lead to frustration such as a lack of time and proper peers to learn together with. Some of this frustration stems from encountering persistent errors or conceptual impasses during coding \cite{bd15}. Furthermore, factors outside the code itself, such as poorly designed interfaces or confusing lesson plans add an unnecessary mental burden for students, which is a strong predictor of frustration \cite{nml23}.

This issue has evolved with modern tools; for instance, recent behavioral studies show that novice programmers spend significant time and cognitive effort in verifying and debugging code generated by AI assistants like Github Copilot, which can itself be a new source of frustration \cite{mbfh24}.

Crucially, these feelings have significant consequences: unresolved frustration can lead to boredom and disengagement, which are directly linked to poorer performance and learning outcomes. Therefore, designing learning environments that can recognize and mitigate these negative emotional states is essential to support novice programmers and improve their chances of success.

\subsection{Conversational Agents in Programming Education}

The field of computer science remains ever growing, and with it come an influx of ever more students that may end up experiencing these aforementioned challenges and frustrations. With the advent of AI and conversational agents, various chatbots have been made or utilized to assist in programming education.

For example, \citet{gvrv24} performed a case study on students and teachers who have previously used ChatGPT and found that students mostly used it for explaining, generating, and debugging code alongside conceptual understanding of course material. In a certain programming assignment, students treated ChatGPT as a tutor to explain code rather than as a tool to generate code, contrary to the expectations of their instructor. \citet{gvrv24} found no connection between AI use and student performance, but do point out that the two highest scorers did not use ChatGPT and admit that the low sample size may be unsuitable for extracting such connections. Although students reported that ChatGPT improved their learning, their instructor lamented that their output quality and perceived learning was lower than that of previous year and suggested that ChatGPT could have decreased student's time investment in the assignment and thus lessened their programming experience.

CS50, an introductory programming course by Harvard University, developed their own ``duck debugger'' to approximate a 1:1 teacher-to-student ratio \cite{lzxpm25}. They found that simply using LLMs based on Generative Pre-trained Transformers (GPT) may circumvent the learning process as they are too helpful and may outright give solutions rather than act as an instructor -- even if prompted to do otherwise. To this end, \citet{lzxpm25} proposed various solutions such as integrating human feedback and providing sample interactions in the system prompt and in a curated dataset which would be used to align the duck debugger towards the pedagogical goals of the course. These solutions demonstrated improved model behavior and alignment with the aforementioned pedagogical goals.

\citet{phgi21} considered responsiveness, scalability, and coverage as key issues that motivated their development of an interactive chatbot for an online Python course. In a similar vein, \citet{h19}'s ``Coding Tutor'' was also successfully developed to ``take over tasks of teaching assistants in times when no human teaching assistant or lecturer is available for help, e.g. due to resource constraints.'' \citet{csmo22}, who developed ``Pyo'', a chatbot that was also meant to facilitate personalized assistance in an online introductory programming class. Pyo provided exercise assistance, error guidance, and concept definitions. Meanwhile, \citet{phgi21} specifically targeted error guidance, developing ``Er-Bot'', a chatbot that responds to error messages with targeted advice; and cited previous research on enhanced compiler error message generation that gave programmers advice when encountering runtime errors in Python code.

Given the wealth of literature, it is evidently clear that there is significant interest in augmenting programming education with conversational agents so that every individual student may receive a personalized education in an era where students vastly outnumber their instructors, teaching assistants, and teaching fellows. There are various weaknesses that have to be accounted for, such as overly-helpful responses and misalignment with pedagogical goals, but there have also been multiple approaches to solve these and to implement these conversational agents.

\subsection{Empathic Approaches and Frameworks}

In an educational context, empathy is essential for fostering positive improvements in student motivation and performance \cite{gsb16,gh14}. As a concept, it involves understanding another's mental state (cognitive empathy), and being affected by it (affective empathy) \cite{z22}. Translating these pedagogical principles into technology is the focus of empathic conversational systems, a field of AI research dedicated to building chatbots that can recognize and respond appropriately to user emotions \cite{ry22}.

While the potential for AI in education is vast, a key challenge is bridging the gap between a system's technical capabilities and the nuanced emotional intelligence of a human educator \cite{ct23}, especially as current tools can foster over-reliance by providing direct solutions rather than teaching users step-by-step \cite{ggkmsmc25,kcpntj25}.

In the context of debugging, the goal of an empathic approach is not just to provide emotional support, but also to mitigate the negative effects of frustration on cognitive performance, thereby helping the student regain the mental clarity needed to solve the technical problem at hand. In fact, recent empirical studies have validated this approach: \citet{opadm24} found that an empathic chatbot providing affective feedback significantly improved student's motivation and self-regulation while they were developing computer competencies.

Knowing this, studies explored the possibility of modeling and improving upon the empathic aspect of artificial intelligence. \citet{cpkahbnb13} outlined key points that contribute to the success of empathic tutors; for example, the understanding of the learner's affective state, applying and adjusting accordingly and striving to form a social bond with the learner. Moreover, \citet{slmtpc25} stated that student or learner engagement is improved through the integration of empathic responses or approaches through emotional recognition and corresponding adaptation.

Recent work in this field has produced advanced technical frameworks like AIVA, which integrates multimodal emotion detection with LLM-driven responses to create a general-purpose virtual companion \cite{l25}. This demonstrates the technical feasibility of our approach, which aims to apply similar principles specifically to the domain of novice programmer debugging.

Across related domains, evidence suggests modest affective benefits: empathic agents reduce frustration \cite{h06}, pedagogical agents slightly lower cognitive load \cite{lwdzw25}, and adaptive feedback increases motivation and persistence \cite{mgfpb20}. While individually limited (they have small sample sizes), these collectively point to the potential of empathic, adaptive systems to sustain learner engagement.

\subsection{Gaps in Knowledge and Research Directions}

Despite its potential, empathy can have counterproductive effects if not carefully balanced. Empathy is not always a positive force in teaching, as it can be susceptible to bias and may even lead to counterproductive learning patterns. For example, students that are praised by tutors for their successes may be led to choose easier tasks as opposed to learning newer and more complex topics in fear of losing that praise \cite{kd99,md98}.

Research exploring the link between teacher empathy and student outcomes, such as classroom behavior and academic performance, has yielded mixed results. A major reason for this inconsistency is the lack of standardized assessment criteria across many of these studies, which casts doubt on the reliability of their findings \cite{ack22}.

While the literature shows a clear trend towards using conversational agents in programming education and a growing interest in developing empathic AI tutors, a specific gap exists at the intersection of these fields. Current research often focuses on conceptual help or general tutoring, with less emphasis on mitigating the in-the-moment frustration a novice programmer would feel when facing a cryptic compiler error. Furthermore, while studies explore empathic responses, few have investigated the use of visual, non-verbal cues, such as an animated agent with different emotional poses, to deliver this feedback in a debugging context.

Therefore, our project aims to address this gap by designing and evaluating a system that provides empathic feedback specifically tailored to GCC compiler errors. This study focuses on CCPROG1 (Logic Formulation and Introductory Programming), the foundational introductory programming course required for all computer science majors at De La Salle University, Manila, Philippines. This course's reliance on the C language and the GCC compiler, known for its cryptic error messages, makes it an ideal context to test our primary goal: reducing frustration and improving the learning experience for novice programmers.

\section{Methodology}

\begin{figure}
	\centering
	\includegraphics[width=\linewidth]{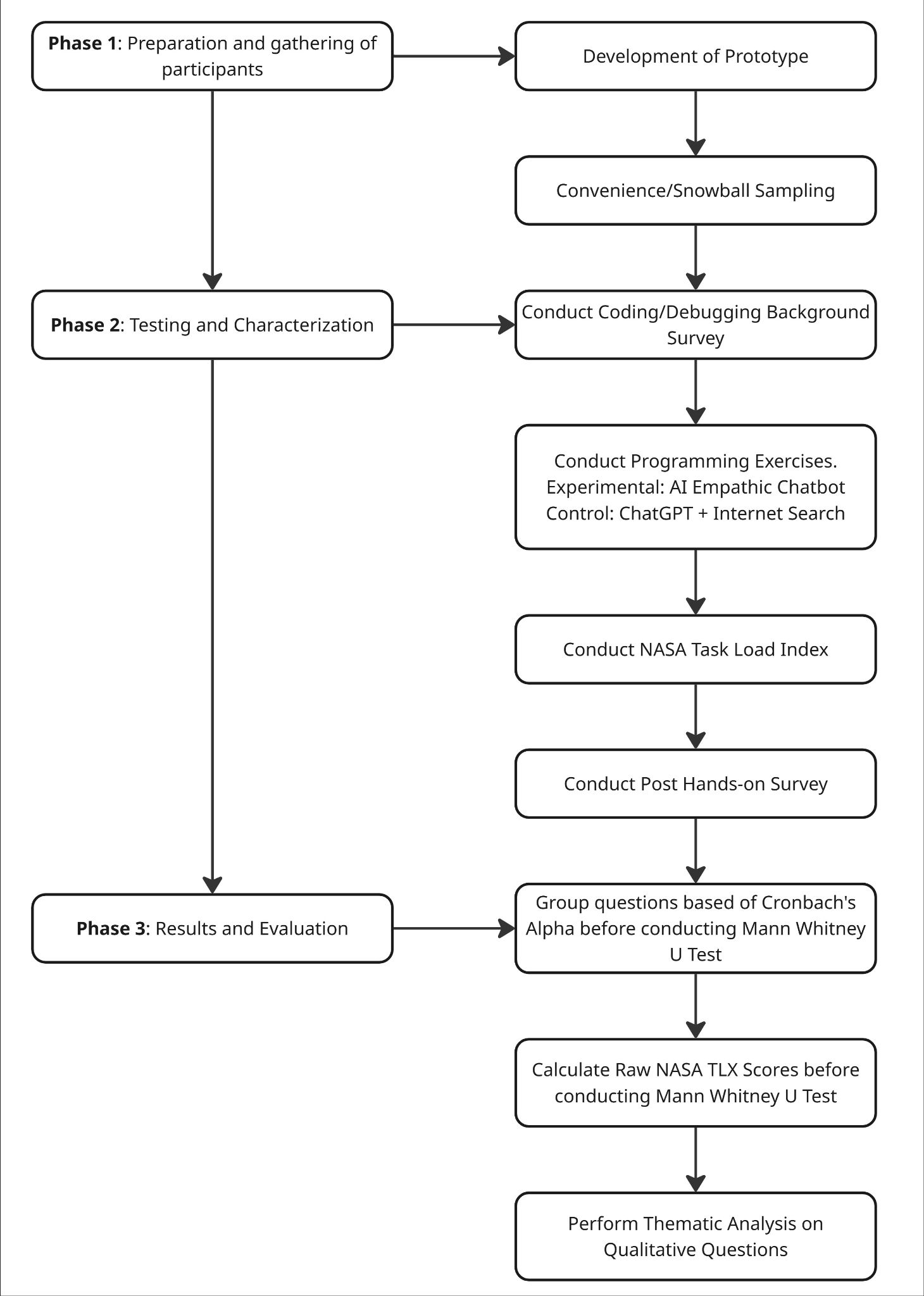}
	\caption{Schematic Diagram of Methodology}
	\label{fig:methodology_diagram}
\end{figure}

This study employs a structured experimental methodology designed to address the research questions stated.
The experiment will be done by dividing the procedure into three phases as seen in the Schematic Diagram shown in \figurename
\ref{fig:methodology_diagram}.

\subsection{Preparation and Participant Plan}

\begin{figure}
	\centering
	\includegraphics[width=\linewidth]{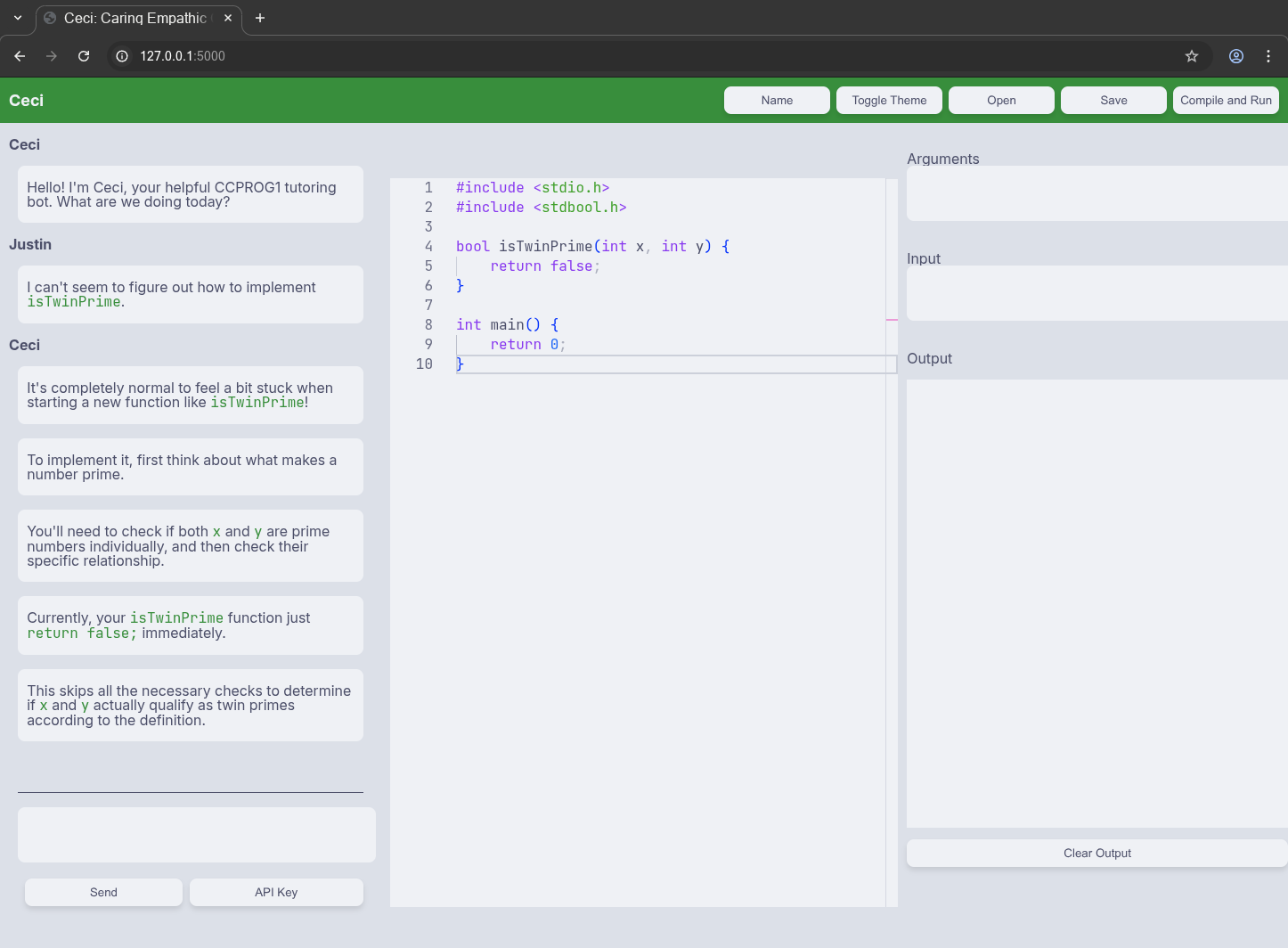}
	\caption{Prototype Interface Design}
	\label{fig:prototype_interface}
\end{figure}

The first step in the methodology is the preparation of necessary components for the study such as the prototype itself and the participants.

\subsubsection{Prototype Design}

The prototype, named Ceci, is an Integrated Development Environment (IDE) run locally through Python Flask.
Within the IDE there would be three sections: the AI chatbot interface, the coding interface, and the terminal as shown in \figurename~\ref{fig:prototype_interface}.

In this interface, the user may chat to prompt the chatbot as well that will give the appropriate empathic response.
The design is somewhat inspired from the study of \citet{h19}, where they developed a coding tutor with a similar interface.
Aside from a standard chat interface, code compilation also triggers the chatbot to respond to the result of the compilation.

The AI chatbot was designed to query Gemini AI through a specific and carefully engineered system prompt to constrain the chatbot behavior to reflect correctness and empathy. To ensure reproducibility and clarity, the system prompt was structured using explicit rule-based constraints. Specifically, Ceci was instructed to:

\begin{itemize}
\item avoid directly providing complete code solutions,
\item prioritize step-by-step guidance through hints,
\item include an affective acknowledgement before technical explanation, and
\item limit responses to short, structured outputs consisting of three parts: emotional acknowledgement, guided suggestion, and technical explanation.
\end{itemize}

For example, when encountering a compilation error, Ceci first acknowledges potential frustration (e.g., ``In C, \texttt{for} loops have a very specific structure.''), followed by a hint (e.g., ``Take a close look at how the different parts inside the parentheses are separated.''), and then a concise explanation of the compiler message (e.g. ``the error `expected ``;'' before ``i''' indicates that C expects a semicolon after the loop's condition (\texttt{i <= n}) before it sees the increment part (\texttt{i++})'').

These constraints were implemented entirely through prompt engineering rather than fine-tuning, ensuring controllability but also limiting adaptability.

\subsubsection{Sampling Method}

The criteria for participation of the study requires the samples to be novice programmers; consequently, the researchers will employ a convenience sampling technique targeting De La Salle University students taking CCPROG1.
Additionally, a screening test would be given to gauge and filter out participants with the intention of keeping the sample to be of purely novice programmers.

\subsubsection{Materials}

The participants will be given a programming exercise to work on.
The exercise is a hands-on type activity where they would implement a program from scratch.
The program the participants were tasked to design is a twin prime identifier within a range of integers.

\subsection{Testing and Characterization}

Once the preparation is completed, the participants will proceed to the experimentation phase, during which the designed exercises and assessments will be conducted.

\subsubsection{Pre-registration}

Participants first pre-register for the experiment through signing up via Google Forms.
Aside from the participants' contact details, the form also asks participants to rate their own proficiency in the C programming language through a series of Likert scale questions (the aforementioned screening test in Section 3.1.2).

\subsubsection{Experimental Design}

The experiment will employ a between groups design following a two group structure consisting of an experimental group, and control group.
The independent variable from these groups will be the technique each will use to debug, where the experimental group will be designated to the AI chatbot interface developed for this study, the control group will have no custom interface and will code with access to ChatGPT \cite{hhthqm25,ltlcl23}.
To address possible confounding variables, all participants will be given the same programming exercise, and the same physical environment as well.
Besides these, any tests conducted before and after will also be identical.

\subsection{Results and Evaluation}

Following the experimentation proper, the researchers will proceed with data collection and evaluation.

\subsubsection{NASA Task Load Index}

NASA Task Load Index (NASA-TLX) is an assessment tool that measures perceived workload or frustration \cite{hs88}.
The NASA-TLX consists of six subscales that represent the total workload of the participant.
These subscales are Mental Demand, Physical Demand, Temporal Demand, Performance, Effort, and Frustration.
NASA-TLX is widely used in various areas such as medical, engineering and even among human-chatbot interaction \cite{cpfs15,ssp21}.
For the purposes of this study, NASA-TLX will be used to quantify subjective perception of workload among participants debugging with the three techniques on each group.

\subsubsection{Post Hands-on Survey}

A questionnaire will be administered to the participants after their participation in the hands-on section of the study.
The questionnaire will contain both open ended and Likert scale type questions designed to answer the research questions.
The open ended questions will be analyzed through thematic analysis to analyze qualitative data \cite{amniaak25}.
This will also be done using inductive coding where themes naturally emerge from the short essay responses.
The Likert scale questions on the other hand will be analyzed through grouping using Cronbach's alpha coefficient, a formula that measures how reliable a set of questions when grouped together making up an underlying construct \cite{c07}.

The formula for Cronbach's alpha is given by Equation~\ref{eq:cronbach}:
\begin{equation}
	\alpha = \frac{k}{k-1}\left(1-\frac{\sum_{i=1}^k\sigma_{y_i}^2}{\sigma_x^2}\right)\label{eq:cronbach}
\end{equation}
where $k$ is the number of items or questions, $\sigma_{y_i}^2$ is the variance per item and $\sigma_x^2$ is the variance of the total scores \cite{rm17}.
The scores of the participants will be averaged for their average score per category.
Results from this averaging and the NASA-TLX will then be analyzed using Mann-Whitney U Test, a nonparametric test to compare two independent groups \cite{ssk23}.

\subsection{Ethical Considerations}

Prior to participation, a consent form will be obtained from all selected participants following the sampling process.
Each participant will be informed on the purpose, procedures, and scope of the study to ensure their participation.
Participants have the option to withdraw from the study at any stage without penalty or consequence.
Furthermore, all personal and research-related data will be treated with strict confidentiality and used solely for academic purposes.
The study will adhere to the Data Privacy Act of 2012, ensuring that identifiable information remains secure and anonymous throughout the study.


\section{Results and Analysis}

\subsection{Experimentation Proper}

The sampling method was able to secure a total of 11 participants, where they are split to 6 experimental group members and 5 control group members.
The researchers utilized the Gokongwei Room 304A Computer Laboratory in DLSU to conduct the study, where the participants' environment remained consistent throughout the experiment (see Figure~3 and Table~\ref{tab:mann_whitney_tlx}).
The participants were given 30 minutes to attempt to complete the programming exercise.
Directly after the hand-on session, the participants were given the NASA TLX and the questionnaire.

\subsection{NASA TLX}

\begin{figure}
	\centering
	\includegraphics[width=\linewidth]{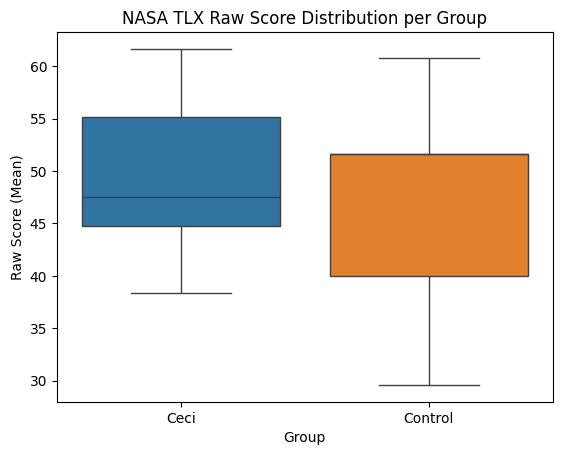}
	\caption{NASA TLX Results per Group. \textnormal{The Ceci group seems to experience a slightly higher task load index than the ChatGPT group, but the difference is not statistically significant as can be seen in \tablename~\ref{tab:mann_whitney_tlx}.}}
	\label{fig:nasa_tlx_results}
\end{figure}

\begin{table}
	\centering
	\caption{Mann-Whitney U Test for NASA TLX Raw Scores}
	\label{tab:mann_whitney_tlx}
	\begin{tabular}{lr}
		\hline
		\textbf{Statistic}            & \textbf{Value}  \\
		\hline
		Significance level ($\alpha$) & 0.05            \\
		Mann-Whitney U                & 16              \\
		\hline
		\textbf{P-Value}              & \textbf{0.9271} \\
		\hline
	\end{tabular}
\end{table}

The results of the NASA TLX are shown in \figurename~\ref{fig:nasa_tlx_results}.
The Raw Score is computed through the arithmetic mean of the individual metrics: The researchers perform a Mann-Whitney U Test Result for Raw Scores per Group, where the null hypothesis states that there is no significant difference between the Ceci and ChatGPT groups while the alternative states that there is a significant difference between the Ceci and ChatGPT groups.
The p-value of 0.9271 is greater than the alpha of 0.05, so the researchers fail to reject the null hypothesis.
That is, there is no significant difference between the task load index of the Ceci and ChatGPT groups.

\subsection{Thematic Analysis}

The researchers performed a thematic analysis to get insight on the respondents answers on the open-ended questions.
Participants A1 to A6 are from the group that used Ceci, while participants B1 to B5 are from the group that used ChatGPT.

\subsubsection{Coping Strategy}

When coping with frustration in programming related tasks, Coping strategies for misunderstanding of a programming concepts among our participants include coping with internet and coping through AI (A2, A3, B4, B5), some use indirect methods of coping such as through taking a break (A3, B5), and some participants cope through tracing through the problem themselves (A3, B5).

Non-AI debugging strategies of the participants include referring to online resources such as documentation (A2, A5, B4), using online forums such as Stack Overflow (B1).
AI strategies are used mostly as a last resort (A1, B2, B4).

\subsubsection{Likes}

Respondents that used ChatGPT had a degree of familiarity when using it to work through programming problems (B1, B2, B3) which provided a degree of comfort and reliability (B2, B3).
Respondents also preferred the instant solutions that ChatGPT provided, and the reliability that comes with knowing that it can provide an answer (B4).
Respondents that used Ceci liked the direct embedding of the AI with the IDE given that it was context aware enough to generate a response on certain actions such as in compiling (A1, A3).
Participants also found Ceci to be friendly and encouraging (A2, A4).

\subsubsection{Dislikes}

Respondents that disliked ChatGPT did not like that they directly copy-pasted some codes to accomplish the task which they viewed as making them overreliant (B1, B5).
Some respondents noted that using ChatGPT was a lot slower than directly integrating and thinking of the code themselves (B2, B3).
Respondents that disliked Ceci did not like some of the UI issues and technical problems such as the need for an API key (A3, A4, A6).
Some respondents also disliked the fact that it wasn't descriptive enough to tackle some problems (A1).

\subsubsection{Preference for Empathicness}

Respondents show mixed feelings with regard to empathic assistance in the context of novice programming work.
Respondents that do not prefer empathic assistance indicate that they prefer more straightforward and logical explanations detached from emotion (A1, A3).
One participant cites that they prefer straightforward answers as they take less time to read (A5).
Respondents that do prefer empathic assistance indicate that empathic responses put them in a good mood (A4) or the adaptive nature of empathic systems to a person who is learning, or who is in a rush or the like (B4).

\subsection{Likert Scale Analysis}

\subsubsection{Grouping through Cronbach's Alpha}

\begin{table*}[ht]
	\centering
	\caption{Reliability Analysis (Cronbach's Alpha)}
	\label{tab:cronbach}
	\begin{tabular}{lccc}
		\hline
		                          & \textbf{Learning} & \textbf{Comfort} & \textbf{Perceived Effectiveness} \\
		\hline
		Items                     & 2, 3, 6           & 5, 8, 9          & 1, 4, 10                         \\
		Sum of Variances per Item & 3.22              & 3.85             & 4.35                             \\
		Number of items           & 3                 & 3                & 3                                \\
		Variance of Total Scores  & 6.36              & 7.40             & 8.25                             \\
		\hline
		\textbf{Cronbach's Alpha} & \textbf{0.741}    & \textbf{0.719}   & \textbf{0.710}                   \\
		\hline
	\end{tabular}
\end{table*}

Using Cronbach's Alpha coefficient, the researchers were able to safely group some questions with a reliability of 0.7 or higher.
The table below, Table~\ref{tab:cronbach} shows the computation and the categorization of the Likert scale questions.
The first set of items is as follows:
\begin{itemize}
	\item ``Programming is more efficient with the setup I was in.''
	\item ``I was able to learn more during the hands-on session.''
	\item ``The tools assigned to me help improve my understanding of the code and concepts.''
\end{itemize}
These items yield an alpha coefficient of 0.741 making it sufficiently reliable to be grouped for a common sentiment like effectivity of learning.

The second set is as follows:
\begin{itemize}
	\item ``The supportive messages helped improve my efficiency.''
	\item ``I feel comfortable and not as stressed with the programming setup I was in.''
	\item ``Debugging and programming this way does not induce that much hassle.''
\end{itemize}
which produced a coefficient of 0.718.
This group then can be categorized as comfort or lack of stress.

The last group had the following items:
\begin{itemize}
	\item ``I had less difficulty programming with my setup.''
	\item ``I heavily relied on the tools provided for me.''
	\item ``I have no problem getting the correct information from the tools I was assigned to.''
\end{itemize}
and have a coefficient of 0.710 that allows the set to be considered as a group, where its category falls more on the perceived effectiveness of the setup.
One question however, question 7 (``The feedback/information I received helped me correct some errors.
'') did not fit reliably within the groups as its inclusion results to a coefficient alpha lesser than 0.7, nullifying the validity of a grouping.
Thus, the researchers have decided to treat this question as a standalone item.

\subsubsection{Mann-Whitney U Test Results}

\begin{table*}
	\centering
	\caption{Mann-Whitney U Test Results for Likert Scale Groups}
	\label{tab:mann_whitney_groups}
	\begin{tabular}{lcccc}
		\hline
		\textbf{Group/Item}      & \textbf{Learning} & \textbf{Comfort} & \textbf{Perceived Effectiveness} & \textbf{Error Correction (Q7)} \\
		\hline
		Significance level       & 0.05              & 0.05             & 0.05                             & 0.05                           \\
		Mann-Whitney U Statistic & 12.5              & 11               & 15.5                             & 25.0                           \\
		\hline
		\textbf{P-Value}         & \textbf{0.7119}   & \textbf{0.5180}  & \textbf{1.0}                     & \textbf{0.0220}                \\
		\hline
	\end{tabular}
\end{table*}

For each group, the answers of the participants are averaged for all questions.
This results in an average score for a given category per participant.
The resulting scores are then analyzed using Mann-Whitney U Test (see Table~\ref{tab:mann_whitney_groups}) to compare the similarity of the Ceci and Control group.
The same is done with item 7, as it did not belong to a group, its analysis is isolated to just itself.
For each group of questions, the p-value is greater than the significance level.
This means that there is no significant difference in learning, comfort, and perceived effectiveness between the Ceci and ChatGPT groups.

The resulting p-value returned from the test is 0.02195 which is less than the significance level of the test.
This suggests that the perceived helpfulness of the Ceci group is significantly greater than that of the control group.

\subsection{Completion Rate}

At the end of their hands-on the participants are tasked to submit their latest code, and they also were required to answer an item in the questionnaire asking if they completed the program or not.
Of the 6 Ceci users, none of them were able to complete while 4 out of 5 of the Control group were able to finish.
This suggests that the current iteration of Ceci may have been detrimental to task completion.
This disparity is most likely attributable to the UI and the technical friction reported by Ceci participants -- particularly the API key requirement flagged in A3, A4, and A6 -- rather than a failure of the empathic approach itself. The control group's prior familiarity with ChatGPT (B1, B2, B3) likely reduced onboarding friction significantly, whereas Ceci users had to contend with a brand-new interface under a strict 30-minute time limit.


\section{Discussion}

\subsection{Effectiveness during coding}

The thematic analysis shows mixed preferences towards the effectiveness of an empathic AI chatbot in coding tasks.
While some appreciated the empathic responses, others felt it was unnecessary towards actually getting the program debugged properly.
These results suggest that learning may depend heavily on an individual level, as some students who prefer to receive encouragement and support may see improvement in using the empathic IDE while the students who prefer direct means of retrieving and obtaining information will benefit less from the empathic responses. This also raises the concern that empathic responses without sufficient technical depth risk feeling unhelpful or even patronizing to task-focused learners. This concern is also supported by participant A1's complaint that Ceci wasn’t descriptive enough to tackle certain problems

These results are also empirically backed up through the Likert scale analysis for the ``Perceived Effectiveness'' group since the outcome of the analysis also indicates no significant difference in perceived effectiveness suggesting that there is no meaningful improvement between having an empathic IDE or not.
However, the results from item 7 suggest that the Ceci group's perceived helpfulness in resolving errors is greater than the Control group, indicating that the application was effective in assisting the students in resolving errors.

\subsection{Extent of learning and understanding}

The result of the test indicates that there is no significant difference between the Ceci and Control groups in learning new concepts and understanding code.
This suggests that, regardless of the presence of an empathic aid, students' perceived learning remains the same.
One possible explanation for this outcome is the mixed reception of empathic responses within the learning environment.
Another factor that may have contributed is the strict time constraint of the hands-on activity; as highlighted in the thematic analysis, some participants preferred direct, straightforward answers that addressed their immediate needs.

\subsection{Effectiveness in reducing frustration}

The NASA-TLX Raw Score Mann-Whitney U Test result indicates that use of Ceci did not significantly affect task load.
The similarity in task load indices for the two groups may be caused by separate problems in the set-up.
For instance, a few responses from the Ceci group regarding what they disliked about the application was that the ``AI is buggy'' or that the ``AI is unhelpful'', while a different issue arose for the control group where they find the need to "to elaborate more or re-paste the code".

Thematic analysis matches the results in the pre-registration survey, with participants that prefer thorough analysis of program and implementation details finding less frustration compared to Ceci purposefully not giving out the full answer.
The ChatGPT group additionally remarked about risks of overreliance on an LLM, which suggests deeper considerations around pedagogical alignment and tool design.

It is also worth noting that comparing Ceci, a first-iteration prototype against VSCode with ChatGPT (A pairing of two mature, widely-used tools) introduced an inherent confounding variable. UI frustrations specific to the prototype, such as the API key requirement and general instability, likely inflated the NASA-TLX frustration scores for the experimental group independently of the empathic AI features themselves. Future studies should ideally conduct usability testing on the prototype prior to comparative experiments, or use a similarly barebones control interface to ensure a fairer comparison.

Another key insight from this study is that empathy alone is insufficient in programming assistance contexts. Participants who preferred direct and concise technical explanations perceived empathic responses as less efficient or even distracting. This suggests that effective AI tutors must balance affective support with domain-specific depth, rather than treating empathy as a standalone feature. This finding highlights a design tension in empathic AI systems: increasing emotional support without sufficient technical grounding may reduce perceived usefulness, particularly for task-oriented users.


\section{Conclusion and Future Work}

\subsection{Conclusion}

The study's results and analysis reveal a nuanced picture. Notably, Ceci users reported significantly greater perceived helpfulness in error correction ($p = 0.0220$), suggesting that an empathic IDE may offer targeted benefits in specific aspects of the debugging experience. However, broader measures of perceived effectiveness, learning, and workload showed no significant difference between the Ceci and Control groups.

This entails that the inclusion of empathic responses does not substantially alter students' perceived effectiveness, learning, or frustration levels during coding tasks.
This may be the case of the specific limitations of the study such as the unforeseen issues in the app implementation, the strict time constraints during the hands-on activity, and the differing individual preferences for either empathic assistance or direct assistance, all of which may have influenced how participants engaged with the system.

Furthermore, the mixed qualitative feedback suggests that empathic features may not provide the same benefits across all learners.
While some students appreciated emotional support, others thought that the empathic layer is unnecessary when they want immediate straightforward responses for coding.
These reactions may imply that empathic design may be more effective when applied to specific subgroups that prefer it.

Despite the absence of significant improvements in overall learning or workload, the findings provide important design implications for future AI-assisted programming environments. Specifically, they highlight the importance of balancing empathy with technical depth, minimizing interface friction in experimental tools, and aligning AI behavior with user expectations for efficiency versus support.
As such, this work contributes not by demonstrating performance gains, but by identifying critical factors that influence the effectiveness of empathic AI in programming education.


\subsection{Future Work}

The study is bound by various time constraints, and future work enjoys a variety of avenues for improvement.
First and foremost, future tests should ensure they test on a bigger sample of students.
This study was only able to accrue around 11 participants as the testing period was near the finals week.
More participants would enhance our analysis, especially our quantitative statistical tests.
Furthermore, most of the criticism against Ceci was directed towards the UI/UX of the IDE.
Improvements on this aspect would likely lead to future participants directing their criticism towards deeper, more technical aspects of the system rather than implementation issues.







\bibliographystyle{ACM-Reference-Format}
\bibliography{bibfile}

\end{document}